\documentclass[a4paper,twocolumn]{article}
\usepackage[T1]{fontenc}
\usepackage[latin9]{inputenc}
\usepackage{array}
\usepackage{verbatim}
\usepackage{float}
\usepackage{url}
\usepackage{graphicx}

\providecommand{\tabularnewline}{\\}
\floatstyle{ruled}
\newfloat{algorithm}{tbp}{loa}
\floatname{algorithm}{Algorithm}

\newenvironment{lyxcode}
{\par\begin{list}{}{
\setlength{\rightmargin}{\leftmargin}
\setlength{\listparindent}{0pt}
\raggedright
\setlength{\itemsep}{0pt}
\setlength{\parsep}{0pt}
\normalfont\ttfamily}%
 \item[]}
{\end{list}}

\usepackage[noend]{algpseudocode}
\begin{document}

\title{Measuring Password Strength: An Empirical Analysis}

\author{Matteo Dell'Amico, Pietro Michiardi and Yves Roudier\\
Institut Eurecom\\
\{matteo.dell-amico, pietro.michiardi, yves.roudier\}@eurecom.fr}
\maketitle
\begin{abstract}
We present an in-depth analysis on the strength of the almost 10,000
passwords from users of an instant messaging server in Italy. We estimate
the strength of those passwords, and compare the effectiveness of
state-of-the-art attack methods such as dictionaries and Markov chain-based
techniques.

We show that the strength of passwords chosen by users varies enormously,
and that the cost of attacks based on password strength grows very
quickly when the attacker wants to obtain a higher success percentage.
In accordance with existing studies we observe that, in the absence
of measures for enforcing password strength, weak passwords are common.
On the other hand we discover that there will always be a subset of
users with extremely strong passwords that are very unlikely to be
broken.

The results of our study will help in evaluating the security of password-based
authentication means, and they provide important insights for inspiring
new and better proactive password checkers and password recovery tools.
\end{abstract}

\section{Introduction}

Even though much has been said about their weaknesses, passwords still
are -- and will be in the foreseeable future -- ubiquitous in computer
authentication systems. A peculiar characteristics of passwords is
that they inherently carry a trade-off between usability and security:
while strong passwords are hard for attackers to guess, they are on
the other hand also difficult for the user to remember. As Richard
Smith paradoxically notes, password best practices imply that {}``the
password must be impossible to remember and never written down''
\cite{Smith2002Strong}. In light of this, it is not very surprising
that users often knowingly choose to use weak passwords or circumvent
security best practices, since they perceive that following them would
get in the way of doing their work \cite{Adams1999Users,Riley2006Password}. 

To think sensibly about the security of systems that use passwords,
it is therefore essential to analyze the characteristics of passwords
chosen by users. In this work, we analyze a large dataset containing
all user passwords from an instant messaging server located in Italy.
Unlike previous empirical studies on passwords \cite{Morris1979Password,Klein1990Foiling,Spafford1992Observing,Narayanan2005Fast,Cazier2006Password,Florencio2007Largescale,Marechal2008Advances},
this paper evaluates the strength of passwords against a variety of
state of the art techniques for candidate generation. The analysis
we conducted benefited from having access to the passwords in unencrypted
form; this made it possible to measure the strength of all of them,
including those that would hardly be cracked even by extremely powerful
attackers.

We evaluate the strength of a password in terms of their associated
search space size, that is the number of attempts that an attacker
would need to correctly guess it. This measure does not depend on
the particular nature of the authentication system nor on the attacker
capabilities: it is only related to the attack technique and to the
way users choose their password. The attack model and the characteristics
of the system will instead define the cost that the attacker has to
pay for each single guess. By combining this cost with our measures
of password strength, it becomes possible to obtain a sound cost-benefit
analysis for attacks based on password guessing on an authentication
system.

As we will show, different attack techniques are advisable depending
on the search space size that the attacker can afford to explore.
This has to be taken into account when proposing and evaluating new
techniques for reducing the search space: they may be effective only
if the strength of the attack falls within a given interval.

We show that password strength has an extremely wide variance: as
a first approximation, the probability to guess a password at each
attempt decreases roughly exponentially as the size of the explored
search space grows. These diminishing returns imply that, in most
cases, an attacker would eventually find a point where the cost of
continuing the attack would not be justified by the probability of
success. This study provides figures that can help designers and administrators
in assessing the security of their systems by evaluating where that
point resides.

\section{Related Work}

\label{sec:Related-Work}In this section we provide a short review
of studies about password security, and make the case for the importance
of measuring password strength. Attacks such as phishing or social
engineering, where the user is misled in communicating the password
to the attacker, are unrelated to password strength and therefore
outside the scope of this work.

\paragraph{Pricing Via Processing}

To defend against intruders who repeatedly try password after password
until they obtain access to the system, it is possible to limit the
rate at which the attacker is allowed to try new passwords by requiring
the user to perform an action with a moderate cost. While legitimate
users would need to perform this action only once every time they
try to log on, an attacker would need to repeat this process many
times, resulting in a disproportionate cost that renders the attack
worthless. The following measures belong to this category:
\begin{itemize}
\item CAPTCHAs \cite{VonAhn2004Telling}, which require solving puzzles
that are difficult without human intervention;
\item key strengthening techniques, which require a few seconds of computation
to derive a key from the passwords; this idea first appeared in the
design of the UNIX system in the late '70s \cite{Morris1979Password}.
A modern key strengthening algorithm, where the computation length
is configurable via the choice of a tunable parameter, is PBKDF2 \cite{Kaliski2000RFC}.
\end{itemize}
It is important to note that these techniques impose a trade-off to
legitimate users: if an honest user has to pay a cost $c$, the attacker
must pay at most $c\cdot s$, where $s$ is the strength of the password
in terms of the number of attempts needed to guess it. The measures
obtained in this paper can be used to estimate costs and benefits
of these systems, and thus to properly tune this $c$ parameter.

An alternative approach blocks accounts after a given number of failed
attempts. This response, however, opens the door to denial of service
attacks on user accounts and is ineffective unless the attack is specifically
targeted towards a single user \cite{Pinkas2002Securing}.

\paragraph{Offline Attacks}

In most cases, the authentication server does not store passwords
in plain text. Instead, it keeps an {}``encrypted'' version of them
which is conceptually analogous to a hash: when a user attempts to
log on, the password they provide is encrypted and compared to the
stored value. In this way, even if an attacker obtains the encrypted
passwords, these cannot be used right away to log on to the system.
To make it costly for the attacker to guess the password by encrypting
lots of password candidates, key strengthening techniques are again
applied. Attacks based on pre-computing the encrypted version of the
most likely passwords \cite{Narayanan2005Fast,Oechslin2003Making}
are defeated with the simple technique of {}``salting'', also known
since the early days of UNIX: that technique works by appending a
random number to the password before encrypting it, and then storing
this number along with the encrypted password.

Since these techniques are based on the idea of making guessing attacks
costly, the password strength that we are measuring is also a key
parameter when evaluating the resilience of a password system to offline
attacks.

\paragraph{Password Recovery}

We measure password strength by taking into account attempts to break
them with state of the art techniques. The free password recovery
software \emph{John the Ripper}%
\footnote{\url{http://www.openwall.com/john/}%
} identifies passwords by checking them against a large-sized dictionary,
plus a fixed set of {}``mangling'' rules, such as appending or prepending
digits to dictionary words. According to Bruce Schneier's description
\cite{Schneier2007Schneier}, AccessData's proprietary Password Recovery
Toolkit complements this approach with a {}``phonetic pattern''
set generated via a Markov chain routine to generate meaningless but
pronounceable passwords. In Section \ref{sec:Markov-Chain-Based-Attack},
we formalize a method based on the same idea and evaluate its merits
in reducing the search space for cracking passwords.

\paragraph{Proactive Password Checking}

A proactive password checker is a system that forces (or advises)
the user to choose complex enough passwords.

The impact of these checkers on actual password security is debatable:
as Wu \cite{Wu1999RealWorld} notes, {}``{[}users are{]} very good
at selecting passwords that are just `good enough' to pass whatever
checking is in place''. The MySpace social network requires users
to have at least a non-alphabetic character in their password; in
a set of leaked passwords, 86\% of the users complied with this requirement
by appending a number at the end of their password; for 20\% of them
that number was a {}``1'' \cite{Porst2007Brief}. Furthermore, a
proactive password checker could encourage users to use non-dictionary
passwords that are related to their personal life such as dates, telephone
numbers or license plate numbers \cite{Adams1999Users}. For a motivated
attacker, these passwords are even easier to guess than dictionary
words. Moreover, a {}``strong password'' in the abstract could force
the user to write it down and leave it in a place where an attacker
can easily find it. For example, many employees hide passwords under
their mouse pads at their companies \cite{Smith2002Strong}.

In general, it seems that password strength checkers actually increase
system security only if they are seen by users as a tool that helps
them and not just as an additional hoop they have to jump through
to get their job done.

Existing password checkers are based on quite naive metrics \cite{Bishop1995Improving,Yan2001Note}:
they check on password length, or resilience to {}``brute force''
and dictionary based attacks; still, they do not take into account
advanced cracking techniques. Our measure of strength as search space
size can be used as the basis for more effective password checkers.

\paragraph{Empirical Studies}

It is a well known fact that many users almost invariably choose easy
to guess passwords; current empirical studies, however, generally
focus on a single kind of attack and neglect to quantify how strong
the remaining share of passwords are with respect to more general
attacks. To the best of our knowledge, no other work evaluates the
strength of passwords over their whole strength spectrum and against
all state-of-the-art techniques.

Analyses on dictionary attacks report a percentage of broken passwords
varying between 17\% and 24\% \cite{Morris1979Password,Klein1990Foiling,Spafford1992Observing}.
In Section \ref{sec:Dictionary-Attack}, before investigating the
remaining stronger passwords, we obtain results of similar magnitude,
varying with the type and size of dictionary used.

Some studies are based on a dataset of encrypted passwords, and only
report on the ones that have been actually cracked \cite{Klein1990Foiling,Narayanan2005Fast,Cazier2006Password,Marechal2008Advances};
in comparison, we had access to the plain-text which gave us information
on the passwords that would be computationally impractical to break.

In a 2007 study \cite{Florencio2007Largescale}, Florencio and Herley
obtained data about the passwords of about 500,000 users. That work
provides interesting insights about user habits, but only quantifies
password strength with a simple {}``bit strength'' measure based
on their length and on the use of uppercase, numeric, and non-alphanumeric
characters; resilience against advanced password-cracking techniques
is not taken into consideration.

\section{Our Dataset}

\label{sec:Our-Dataset}Our dataset contains the unencrypted passwords
for the 9,317 registered users of an Italian instant messaging server.
Storing passwords in plain text on the server is required by authentication
algorithms such as CRAM-MD5%
\footnote{\url{http://tools.ietf.org/html/rfc2195}%
}. User registration is free and no policy for password strength is
enforced: even the empty password is allowed. The absence of strength
enforcement allows us to investigate the behavior of users when choosing
their password in the absence of external requirements.

Users are free to choose any unused username when registering. A total
of 269 users (2.89\% of the total) use the same string as both username
and password. The single most effective attempt to guess a given user's
password would therefore be its own username.

Some users share the same password, and this results in 7,848 unique
passwords. While in some cases this may be due to coincidences and
use of too frequent passwords, other cases may be the consequence
of the same people registering under different usernames at the same
server.

\begin{figure}
\begin{centering}
\includegraphics[width=0.9\columnwidth]{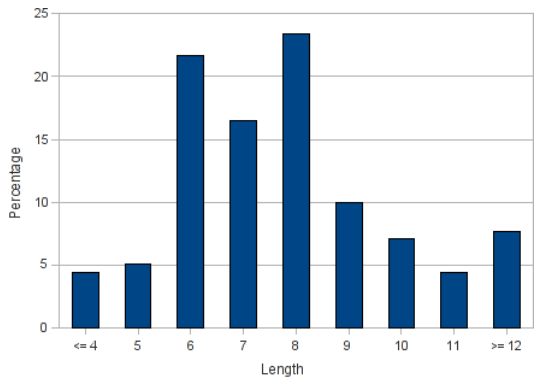}
\par\end{centering}

\caption{\label{fig:Password-length-distribution}Password length distribution.}

\end{figure}

The average password length is 7.86. Figure \ref{fig:Password-length-distribution}
shows the length distribution. Even though the full Unicode character
set is usable for the passwords, only 124 different characters had
been used. Frequencies of characters have very uneven distributions
(see table \ref{tab:Character-distribution}): while one character
out of 11 is an `\texttt{a}', the most frequent uppercase character
(`\texttt{A}') has a frequency of approximately 1 in 500.

%
{}

\begin{table}
\begin{centering}
\begin{tabular}{|c|c|c|}
\hline 
Character & Count & Percentage\tabularnewline
\hline
\hline 
\texttt{a} & 6,681 & 9.12\%\tabularnewline
\hline 
\texttt{e} & 4,520 & 6.17\%\tabularnewline
\hline 
\texttt{o} & 4,484 & 6.12\%\tabularnewline
\hline 
\texttt{i} & 4,388 & 5.99\%\tabularnewline
\hline 
\texttt{r} & 3,628 & 4.95\%\tabularnewline
\hline 
\texttt{n} & 3,310 & 4.52\%\tabularnewline
\hline 
\texttt{l} & 3,095 & 4.23\%\tabularnewline
\hline 
\texttt{s} & 2,895 & 3.95\%\tabularnewline
\hline 
\texttt{t} & 2,853 & 3.90\%\tabularnewline
\hline 
\texttt{1} & 2,518 & 3.44\%\tabularnewline
\hline 
\texttt{c} & 2,367 & 3.23\%\tabularnewline
\hline 
\texttt{m} & 2,137 & 2.92\%\tabularnewline
\hline 
\texttt{0} & 1,990 & 2.72\%\tabularnewline
\hline 
\texttt{p} & 1,945 & 2.66\%\tabularnewline
\hline 
\texttt{d} & 1,813 & 2.48\%\tabularnewline
\hline 
\texttt{2} & 1,692 & 2.31\%\tabularnewline
\hline 
\texttt{u} & 1,640 & 2.40\%\tabularnewline
\hline 
\texttt{b} & 1,624 & 2.22\%\tabularnewline
\hline 
\texttt{3} & 1,487 & 2.03\%\tabularnewline
\hline 
\texttt{g} & 1,334 & 1.82\%\tabularnewline
\hline 
other & 16,832 & 22.98\%\tabularnewline
\hline
\end{tabular}
\par\end{centering}

\caption{\label{tab:Character-distribution}Character distribution.}

\end{table}

In table \ref{tab:Regular-expressions.}, we show the matching ratio
of various simple regular expression. More than 50\% of the passwords
contain only lowercase characters, and less than 7\% contain non-alphanumeric
characters. Around 15\% of them consist of a string of lowercase characters
followed by a numeric appendage.

\begin{table}
\begin{centering}
\begin{tabular}{|c|c|c|}
\hline 
Expression & Example & Matches\tabularnewline
\hline
\hline 
\texttt{{[}a-z{]}+} & \texttt{abcdef} & 51.20\%\tabularnewline
\hline 
\texttt{{[}A-Z{]}+} & \texttt{ABCDEF} & 0.29\%\tabularnewline
\hline 
\texttt{{[}A-Za-z{]}+} & \texttt{AbCdEf} & 53.74\%\tabularnewline
\hline 
\texttt{{[}0-9{]}+} & \texttt{123456} & 9.10\%\tabularnewline
\hline 
\texttt{{[}a-zA-Z0-9{]}+} & \texttt{A1b2C3} & 93.43\%\tabularnewline
\hline 
\texttt{{[}a-z{]}+{[}0-9{]}+} & \texttt{abc123} & 14.51\%\tabularnewline
\hline 
\texttt{{[}a-zA-Z{]}+{[}0-9{]}+} & \texttt{aBc123} & 16.30\%\tabularnewline
\hline 
\texttt{{[}0-9{]}+{[}a-zA-Z{]}+} & \texttt{123aBc} & 1.80\%\tabularnewline
\hline 
\texttt{{[}0-9{]}+{[}a-z{]}+} & \texttt{123abc} & 1.65\%\tabularnewline
\hline
\end{tabular}
\par\end{centering}

\caption{\label{tab:Regular-expressions.}Percentage of passwords matching
various regular expressions.}

\end{table}

We also analyzed a set of 33,671 leaked MySpace passwords \cite{Grimes2006MySpace,Porst2007Brief}.
Since these passwords have been obtained through a phishing attack,
they include those of less security-conscious users who fell for the
attack. Moreover, MySpace requires users to insert non-alphabetic
characters in their passwords, and this imposes an artificial impact
on passwords that users, left alone, would choose. For these reasons,
we consider this dataset less representative of actual user passwords
than our primary one; we however use it in this work to corroborate
some of our findings by validating them on another dataset.

\section{Dictionary Attack}

\label{sec:Dictionary-Attack}Dictionary attack is the most effective
technique to guess the weakest passwords. We evaluated password strength
by using the dictionaries available in the already mentioned \emph{John
the Ripper} (JtR) password recovery tool. The extended dictionaries
that we used are available for paid download from the program website%
\footnote{\url{http://www.openwall.com/wordlists/}%
}.

\subsection{The Dictionaries}

The JtR dictionaries contain words from 21 different human languages,
plus a list of frequently used passwords. For some languages (like
English and Italian), various dictionaries of different sizes are
available: the smaller ones contain only the most frequently used
words while the bigger ones also contain more obscure words, the rationale
being that more common words are more likely to be chosen as passwords.
Taken together, all dictionaries account for almost 4 million words.
A bigger dictionary containing more than 40 millions words is obtained
using {}``mangling'' rules that attempt to create more complex passwords
by altering dictionary words, for example by juxtaposition of dictionary
words or by appending a number at the end of the word.

An often-advised technique to create strong but easy to remember passwords
is to turn phrases into passwords by extracting an acronym, possibly
also using punctuation. For example, the phrase {}``Alas, poor Yorick!
I knew him, Horatio'' becomes {}``A,pY!Ikh,H''. We also evaluated
such acronyms with a dictionary created by Kuo et al. \cite{Kuo2006Human}
that was put together by scraping websites displaying memorable phrases,
such as citations and music lyrics.

\subsection{Experimental Results}

We simulated dictionary attacks with all the JtR dictionaries. Table
\ref{tab:Dictionary-attacks} shows the results for the most representative
instances.

\begin{table}
\begin{centering}
\begin{tabular}{|c|c|c|c|}
\hline 
Dictionary & Size & Found & Guess prob.\tabularnewline
\hline
\hline 
Frequent passwords & 3,114 & 7.25\% & $2.33\cdot10^{-5}$\tabularnewline
\hline 
English 1 lc & 27,424 & 4.91\% & $1.79\cdot10^{-6}$\tabularnewline
\hline 
English 2 lc & 296,809 & 9.42\% & $3.17\cdot10^{-7}$\tabularnewline
\hline 
English 3 lc & 390,532 & 11.59\% & $2.97\cdot10^{-7}$\tabularnewline
\hline 
English extra lc & 444,678 & 8.03\% & $1.81\cdot10^{-7}$\tabularnewline
\hline 
Italian 1 lc & 63,041 & 3.71\% & $5.89\cdot10^{-7}$\tabularnewline
\hline 
Italian 2 lc & 344,074 & 14.89\% & $4.33\cdot10^{-7}$\tabularnewline
\hline 
All above & 1,117,767 & 25.51\% & $2.28\cdot10^{-7}$\tabularnewline
\hline 
All JtR dictionaries & 3,917,193 & 25.94\% & $6.62\cdot10^{-8}$\tabularnewline
\hline 
All JtR + mangling & 40,532,676 & 30.12\% & $7.43\cdot10^{-9}$\tabularnewline
\hline 
Mnemonics \cite{Kuo2006Human} & 406,430 & 1.27\% & $3.12\cdot10^{-8}$\tabularnewline
\hline
\end{tabular}
\par\end{centering}

\caption{\label{tab:Dictionary-attacks}Dictionary attacks. The {}``lc''
acronym stands for all-lowercase dictionaries: those containing uppercase
letters are matched by very few words in our dataset. The English
1, English 2 and English 3 dictionaries, like Italian 1 and Italian
2, are listed in growing size; each word belonging to a smaller dictionary
is also contained in the bigger versions.}

\end{table}

The {}``found'' column lists the percentage of passwords that appear
in that dictionary; the {}``guess probability'' column reflects
the probability that a random word from that dictionary matches a
random password: a rational attacker would try a word from that dictionary
only if the benefit of cracking the password exceeds the inverse of
that probability times the cost of the effort for trying that password.

The {}``English extra'' dictionary has a slightly misleading name:
it contains words that don't appear in a regular dictionary but that
users are likely to use, such as proper nouns, common misspellings
or alterations of words. Many of them are language-agnostic (e.g.,
{}``Aldebaran'') or come from non-English languages ({}``Mariela'').

As the server is in Italy, most users are Italian. The amount of English
words found in passwords is not particularly surprising for those
who know the tendency that natives have towards the heavy use (and
abuse) of English. An interesting feature is the noticeably higher
density of common English words (those present in the small {}``English
1'' dictionary); that phenomenon is much less relevant with respect
to Italian. We think that this is caused by the fact that most users
know English as a second language, and thus are less inclined to use
an obscure word as their password. This suggests that it might be
good practice to use one's native language to create stronger passwords.

The most important lesson drawn from this data is the principle of
\emph{diminishing returns}: the probability of guessing a word sharply
decreases as the dictionary grows. The 3,100-word dictionary of frequent
passwords cracks 7\% of those in our datasets; by increasing roughly
300 times the size of the dictionary up to more than one million and
including all Italian and English words, the number of cracked passwords
rises to 25\%. When the number of attempts grows beyond 40 millions
by including other languages and mangling, only 5\% more of the passwords
are found. To put it in another way, the probability of guessing a
given password by trying an element of the {}``frequent passwords''
dictionary is one in 43,000. On the other hand, after having tried
all the frequent passwords and the Italian and English dictionary,
the probability of guessing by using another dictionary word is less
than one in 500 million! Since the guessing probability decreases
so sharply, it is conceivable that in many cases it won't be worth
trying a bigger dictionary for the attacker.

We also observe that the mnemonic dictionary is quite ineffective.
This may be due to several reasons: first, few users actually use
mnemonics for their passwords; second, they are actually much harder
to break with dictionary attacks. Moreover, we are not able to ascertain
whether the habit of choosing English passwords for Italian users
would carry over to the use of mnemonics. Our data is, at the moment,
insufficient to point towards one reason or the other.

\section{Markov Chain-Based Attack}

\label{sec:Markov-Chain-Based-Attack}The fact that dictionaries fall
short does not mean that an attacker would need to resort to an exhaustive
brute-force attack: some passwords are much more likely to be chosen
than others. As seen in Section \ref{sec:Our-Dataset}, there is a
very uneven distribution of character choice. Moreover, other regularities
exist: passwords are usually made of pronounceable sub-strings and/or
sequences of keys that are close on the keyboard.

In this section, we describe and validate an attack based on Markov
chain-based modeling of the frequencies of sub-strings with parametric
length $k$, or $k$-graphs. This allows us to label candidate passwords
with variable probabilities, where strings that are labeled as more
likely are checked first. Some password generating utilities actually
use this kind of modeling to obtain meaningless but pronounceable
passwords on the grounds that they're easier to remember, thus sacrificing
some strength for usability%
\footnote{See for example \texttt{gpw} (\url{http://www.multicians.org/thvv/tvvtools.html#gpw}),
\texttt{apg} (\url{http://www.adel.nursat.kz/apg/}), \texttt{otp}
(\url{http://www.fourmilab.ch/onetime/}).%
}.

\subsection{The Technique}

We base our formalization on the techniques shown in \cite{Narayanan2005Fast},
extending the model so that it applies to sub-strings of length 3
and more. This model represents a password choice as a sequence of
random events: first, the length of the password is chosen according
to a given probability distribution; then, each character of the string
gets extracted according to a conditional probability depending on
the previous $k-1$ characters.

We encode the characteristics of passwords via two functions, $\lambda$
and $\nu$. $\lambda$ represents the length distribution of passwords
so that, for example, $\lambda\left(8\right)$ is the probability
that the password has length 8. $\nu$, instead, represents the conditional
probability of each $k$-graph with respect to the corresponding $\left(k-1\right)$-graph:
$\nu\left(c_{1}\ldots c_{k}\left|c_{1}\ldots c_{k-1}\right.\right)$
is the probability that the character $c_{k}$ follows the sub-string
$c_{1}\ldots c_{k-1}$. For $k=1$, $\nu\left(c\right)$ expresses
the frequency of $c$, that is, the probability that a random character
in a password coincides with $c$.

By choosing $k=1$, thus focusing on character frequency, the probability
$P_{1}\left(\alpha\right)$ that our model will generate a string
$\alpha$ (where its length is $\left|\alpha\right|$ and its $i$th
character is $\alpha_{i}$) is

\[
P_{1}\left(\alpha\right)=\lambda\left(\left|\alpha\right|\right)\prod_{1\leq i\leq\left|\alpha\right|}\nu\left(\alpha_{i}\right).\]

To derive $P_{k}$ with $k\geq2$, we will adopt the convention that
$\alpha_{i}={\normalcolor \bot}$ whenever $i\leq0$, where {}``$\bot$''
is a special character not allowed to appear in passwords. For example,
we write the probability that a password starts with the {}``$\mathtt{a}$''
character as $\nu\left(\mathnormal{"\mathtt{\bot a}"}\left|\mathnormal{"\mathtt{\bot}"}\right.\right)$;
the probability that a {}``$\mathtt{b}$'' follows an initial {}``$\mathtt{a}$''
is instead $\nu\left(\mathnormal{"\mathtt{\bot ab}"}\left|\mathnormal{"\mathtt{\bot a}"}\right.\right)$.
Given this, we can formalize the digraph-based probability $P_{2}$
as \[
P_{2}\left(\alpha\right)=\lambda\left(\left|\alpha\right|\right)\prod_{1\leq i\leq\left|\alpha\right|}\nu\left(\alpha_{i-1}\alpha_{i}\left|\alpha_{i-1}\right.\right)\]
and, in general, 

\[
P_{k}\left(\alpha\right)=\lambda\left(\left|\alpha\right|\right)\prod_{1\leq i\leq\left|\alpha\right|}\nu\left(\alpha_{i-k+1}\ldots\alpha_{i}\left|\alpha_{i-k+1}\ldots\alpha_{i-1}\right.\right).\]

\subsubsection{Estimating $\nu$ and $\lambda$}

It is obviously important that the probabilities encoded in the $\lambda$
and $\nu$ functions are representative of the real characteristics
of passwords. We do this by adopting a set of strings as a training
set and setting $\lambda\left(x\right)$ as the fraction of strings
of length $x$. Denoting $C$ as the character set and $\sigma\left(c_{1}\ldots c_{k}\right)$
as the number of occurrences of the sub-string $c_{1}\ldots c_{k}$
in the whole training set, we set\[
\nu\left(c_{1}\ldots c_{k}\left|c_{1}\ldots c_{k-1}\right.\right)=\frac{\sigma\left(c_{1}\ldots c_{k}\right)}{\sum_{\overline{c}\in C}\sigma\left(c_{1}\ldots c_{k-1}\overline{c}\right)}.\]

In the absence of a representative training set of passwords, a dictionary
can be used as in \cite{Narayanan2005Fast}. As we will experimentally
show in Section \ref{sub:Experimental-results}, using passwords themselves
as training set finally results in a better model. In this case, when
computing $P_{k}\left(\alpha\right)$ in our experiments, $\alpha$
itself must be removed from the training set and should not be taken
into account when computing the values of $\lambda$ and $\sigma$.

As mentioned in Section \ref{sec:Our-Dataset}, some users share the
same password. This might be due to chance and to the fact that those
passwords are quite trivial; another possibility is that they come
from the same user registering many accounts and using the same passwords
for all of them. In the latter case, an attacker would not have access
to the password in a representative training set, and it would be
correct for our purposes to remove all copies of the password from
the training set. Since we cannot discriminate between the two cases,
we will adopt a conservative approach that may result in overestimating
the capabilities of the attacker, therefore discarding only a single
copy of the password from the training set.

A model with higher values of $k$ should be more accurate, but the
process of creating it is more difficult and expensive. In the extreme,
a model with $k$ exceeding the maximum password length would explicitly
list the probability of occurrence of each possible password: this
would require prohibitive training set size and storage capabilities
(the required space is of the order of $\left|C\right|^{k}$, where
$\left|C\right|$ is the size of the character set). With limited
resources, when a $k$-graph does not appear in the training set due
to under-sampling, then the probability of a password containing that
$k$-graph is computed as 0. Such a model would therefore never generate
the required password.

\subsubsection{Computing The Search Space Size}

So far, we have described a model that assigns probabilities to passwords,
with the aim of measuring how likely it is that a user would actually
select a given password. A rational attacker would use this model
by enumerating candidate passwords starting with the most likely ones
and continuing in decreasing order of probability.

In order to measure the search space size that such a strategy would
need to explore before finding a given password, we have to find out
how many unsuccessful candidates would be generated before the correct
one: if the Markovian model labels the probability of a password as
$p$, its associated search space size would therefore be the number
of strings with probability of occurrence higher than or equal to
$p.$

\paragraph*{Explicit Counting}

The most obvious system for computing the search space size up to
a given threshold is to plainly enumerate it. In Algorithm \ref{alg:Explicit-counting-of},
we show how this can be implemented with a simple recursive algorithm.

\begin{algorithm}
\begin{algorithmic}
\Function{size}{$c_1\ldots c_{k-1}, l, t$}
\State \Comment{$c_1\ldots c_{k-1}$\textrm{: state, $l$\textrm{: string length}, $t$\textrm{: threshold}}}
\State \textbf{if $l=0$ then return $1$}
\State $s\gets 0$
\ForAll{$\overline c \in C$}
\State $p\gets \nu\left(\left.c_1\ldots c_{k-1} \overline c \right| c_1\ldots c_{k-1}\right)$
\If{$p \geq t$}
\State{$s \gets s + \textsc{size}\left(c_2\ldots c_{k-1} \overline c, l - 1, t \cdot p\right)$}
\EndIf
\EndFor
\State \textbf{return} $s$
\EndFunction
\State
\Function{total\_size}{$t$} \Comment{$t$\textrm{: threshold}}
\State \textbf{return} $\sum_i  \textsc{size}\left(\bot\ldots\bot, i, t \cdot \lambda(i)\right)$
\EndFunction
\end{algorithmic}

\textbf{}%
{}

\caption{\label{alg:Explicit-counting-of}Explicit counting of search space
size.}

\end{algorithm}

\paragraph*{Approximate Estimation}

As the search space grows, the above approach becomes extremely expensive
and should be replaced with an approximate estimation method \cite{Narayanan2005Fast}.
By fixing a base $b>1$, any probability $p$ can be approximate as
$b^{-l}$ for an integer value $l\geq0$. Choosing $l=\left\lfloor -\log_{b}p\right\rfloor $
approximates $p$ by excess, while $l=\left\lceil -\log_{b}p\right\rceil $
approximates by defect. To help intuition, $l$ can be seen as a discrete
{}``password strength'' value, which can be computed as the sum
of strengths for each $k$-graph contained in the password. Values
of $b$ closer to 1 result in a finer granularity for our approximation,
at the cost of an increase in computation.

\begin{algorithm}
\begin{algorithmic}
\Function{appr\_size}{$c_1\ldots c_{k-1}, l, t$}
\State \Comment{$c_1\ldots c_{k-1}$\textrm{: state, $l$\textrm{: string length}, $t$\textrm{: log-threshold}}}
\State \textbf{if $l=0$ then return $1$}
\State $s\gets 0$
\ForAll{$\overline c \in C$}
\State $\overline t \gets t - \left\lceil-\log_b\nu\left(\left.c_1\ldots c_{k-1} \overline c \right| c_1\ldots c_{k-1}\right)\right\rceil$
\If{$\overline t \geq 0$}
\State{$s \gets s + \textsc{cache\_size}\left(c_2\ldots c_{k-1} \overline c, l - 1, \overline t\right)$}
\EndIf
\EndFor
\State \textbf{return} $s$
\EndFunction
\State
\Function{cache\_size}{$c_1\ldots c_{k-1}, l, t$}
\State \Comment{\textrm{We store results from approx\_size in a cache $K$}}
\If{$\left(c_1\ldots c_{k-1}, l, t\right) \notin K$}
\State $K\left(c_1\ldots c_{k-1}, l, t\right) \gets \textsc{appr\_size}\left(c_1\ldots c_{k-1}, l, t\right)$
\EndIf
\State \textbf{return} $K\left(c_1\ldots c_{k-1}, l, t\right)$
\EndFunction
\State
\Function{total\_size}{$t$} \Comment{$t$\textrm{: threshold}}
\State \textbf{return} $\sum_i  \textsc{cache\_size}\left(\bot\ldots\bot, i, \left\lfloor-\log_b t\cdot \lambda(i)\right\rfloor\right)$
\EndFunction
\end{algorithmic}

%
{}

\caption{\label{alg:Implicit-approximation-of}Approximation of search space
size.}

\end{algorithm}

By adopting such an alteration, the computation gets a big speedup
by memoizing the parameters and results of each \textsc{approx\_size}
call, and returning them when the function is called again with the
same parameters. This couldn't be done with the former version, since
the $t$ threshold parameter of the \textsc{size} function is a floating
point number which is very likely to be different at each function
call. 

Since we are aiming for a conservative estimate for the search space
that approximates by excess the capabilities of the attacker, we use
approximations to obtain a lower limit for the search space size.
To do this, we approximate the starting threshold by defect and all
the $\nu$ probabilities by excess.

The result of these modifications is the approximate function defined
in Algorithm \ref{alg:Implicit-approximation-of}.

\subsection{\label{sub:Experimental-results}Experimental results}

This section describes the results of the experiments described above
when applied to our password dataset. Unless otherwise specified,
we use the passwords themselves as training set.

%
{}

\begin{figure}
\begin{centering}
\includegraphics[width=1\columnwidth]{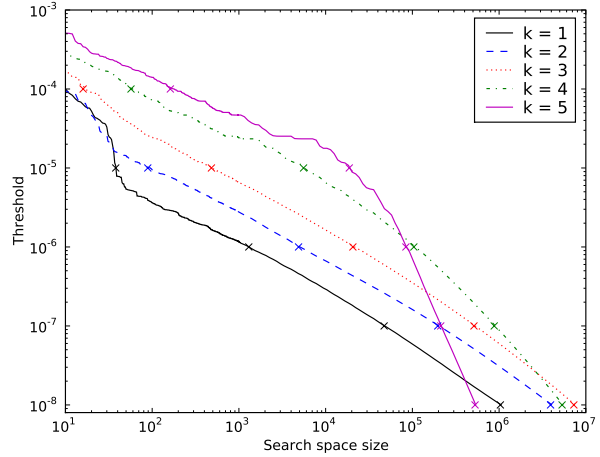}
\par\end{centering}

\caption{\label{fig:Estimation-of-search}Search space size versus probability
threshold for the $k$-graph Markovian model. The plotted curves show
the result of the exact computation of Algorithm \ref{alg:Explicit-counting-of},
while the points marked by crosses are the result of the approximation
of Algorithm \ref{alg:Implicit-approximation-of}.}

\end{figure}

\paragraph{Search Space Size Versus Probability Threshold}

In Figure \ref{fig:Estimation-of-search} we show the size of search
space containing strings labeled with a probability greater or equal
to a given probability threshold. This is computed for different values
of $k$ and using both the exact count and the approximate measure
from Algorithm \ref{alg:Implicit-approximation-of}. We used a parameter
$b=1.01$; with that choice, we obtained a relative error of the order
of 5\% (not noticeable in the figure due to the log-log scale).

By choosing $1\leq k\leq3$ (i.e., basing the model on sub-strings
of lengths 1 to 3), the probabilities of strings generated by the
model roughly follow a power law. It is interesting to note that this
mirrors frequencies of words in human natural languages, which obey
the power law as well \cite{Zipf1949Human}. For $k\geq4$, the number
of candidate strings grows definitely slower as the probability threshold
increases; this is due to the fact that each $k$-graph is represented
by a low number of strings in the training set, and the number of
strings that can be obtained by combining $k$-graphs that are present
in the dataset is limited. We conjecture that, with a bigger training
set, we would obtain a power-law distribution also in this case.

In the following, we use the approximate approach to estimate the
search space size where the exact value becomes either impractical
or impossible to compute. We compute data points for each $p=10^{-i}$
threshold ($i$ being an integer) and interpolate with the power law
that connects the points (a straight line in the log-log plot).

\begin{figure}
\begin{centering}
\includegraphics[width=1\columnwidth]{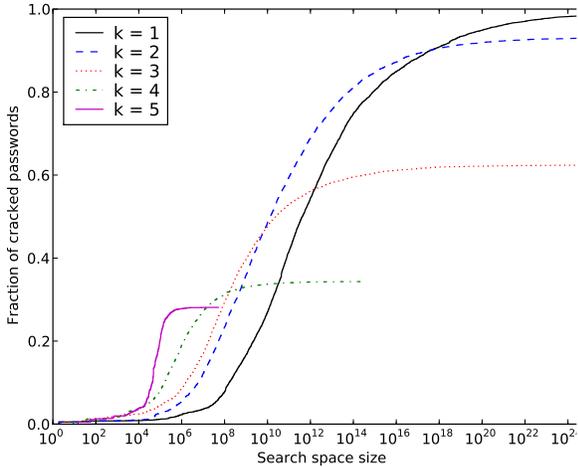}
\par\end{centering}

\caption{\label{fig:Search-space}Search space size versus fraction of guessed
passwords.}

\end{figure}

\paragraph{Password Strength}

In Figure \ref{fig:Search-space}, we plot the fraction of passwords
guessed as a function of the search space size.

With higher values of $k$, we obtain better results for the weaker
passwords due to the more precise modeling obtained in this case.
However, the passwords that include $k$-graphs not represented in
the training set cannot be guessed. Methods based on smaller $k$
values become more effective because they can {}``generalize'' some
more. In practice, the optimal strategy depends on the resources of
the attacker, measured by the number of attempts that can be tried.

The {}``diminishing returns'' effect that we discovered for dictionary
attacks also applies to this technique: even when choosing the best
value of $k$ for each case, around 100,000 candidates need to be
tried in order to guess 20\% of the passwords ($k=5$); this number
rises to roughly 1.1 billions candidates for a success rate of 40\%
($k=3$); the search space needed to break 90\% of the passwords grows
to approximately $3\cdot10^{17}$ ($k=2)$. With such a huge variance
in the size of the search space, it seems that no reasonable attack
based on password guessing would succeed in guessing all passwords
-- excepting those cases where users are artificially forced to limit
password strength, for example by imposing a maximum length.

\paragraph{MySpace Passwords}

\begin{figure}
\begin{centering}
\includegraphics[width=1\columnwidth]{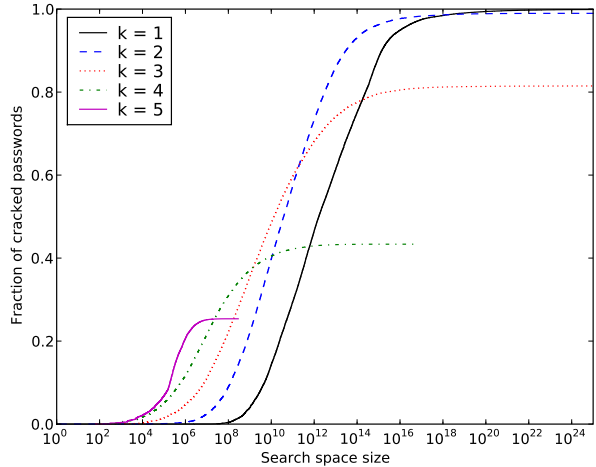}
\par\end{centering}

\caption{\label{fig:Search-space-1}Search space size versus fraction of guessed
passwords on the MySpace dataset.}

\end{figure}

In Figure \ref{fig:Search-space-1}, we repeat our measurements using
MySpace passwords in the place of our main dataset both as training
set and as guessed passwords. We obtain qualitatively similar results
-- in particular, higher values of $k$ are more appropriate as training
sets for weaker passwords, and the diminishing returns principle holds.
From a quantitative point of view, the search space for weak password
is bigger, while it is smaller for stronger passwords. We think that
this is mainly due to the particularities of the dataset: weak passwords
are made stronger by the requirement of non-alphabetic characters;
strong passwords created by security-conscious users, on the other
hand, are under-represented since such users are not likely to fall
victim to a phishing attack.

\paragraph{Brute Force}

\begin{figure}
\begin{centering}
\includegraphics[width=1\columnwidth]{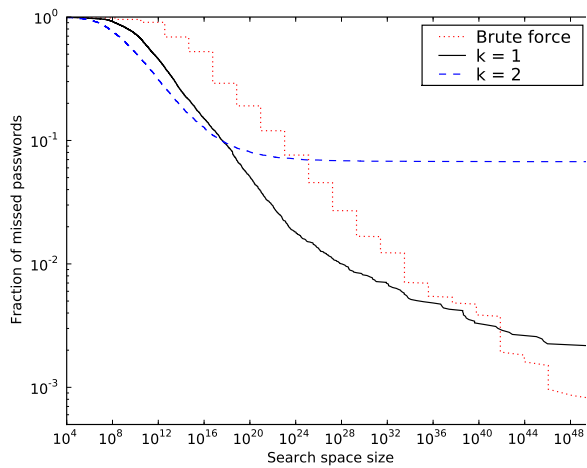}
\par\end{centering}

\caption{\label{fig:Brute-force.}Comparison of brute force and Markov-model
based attacks.}

\end{figure}

In Figure \ref{fig:Brute-force.}, we compare the brute force approach
with our Markovian modeling. The brute force approach starts by trying
the empty password, then proceeds with enumerating all possible passwords
with increasing length. The full Unicode character set currently has
more than 99,000 characters%
\footnote{\url{http://www.unicode.org/press/pr-ucd5.0.html}%
}, but many of them are very rare and definitely unlikely in a password;
to account for this, we again took a conservative approach overestimating
the attacker capabilities, and took into account only the 124 characters
that we have found in our dataset.

In all but the most extreme cases, the Markovian model proves more
efficient by orders of magnitude. It is not before $10^{40}$ candidates
(and having found 99.7\% of the passwords) that a brute force approach
becomes more effective than the Markovian model with $k=1$ (character
frequencies). This number is well beyond the capabilities of any realistic
attacker: to put this in context, a cluster of a thousand 10 GHz machines
would need more than $3\cdot10^{19}$ years to reach that number of
iterations, even assuming that they are able to try a password for
each clock cycle.

\paragraph{Strength and Password Length}

\begin{figure}
\begin{centering}
\includegraphics[width=0.5\columnwidth]{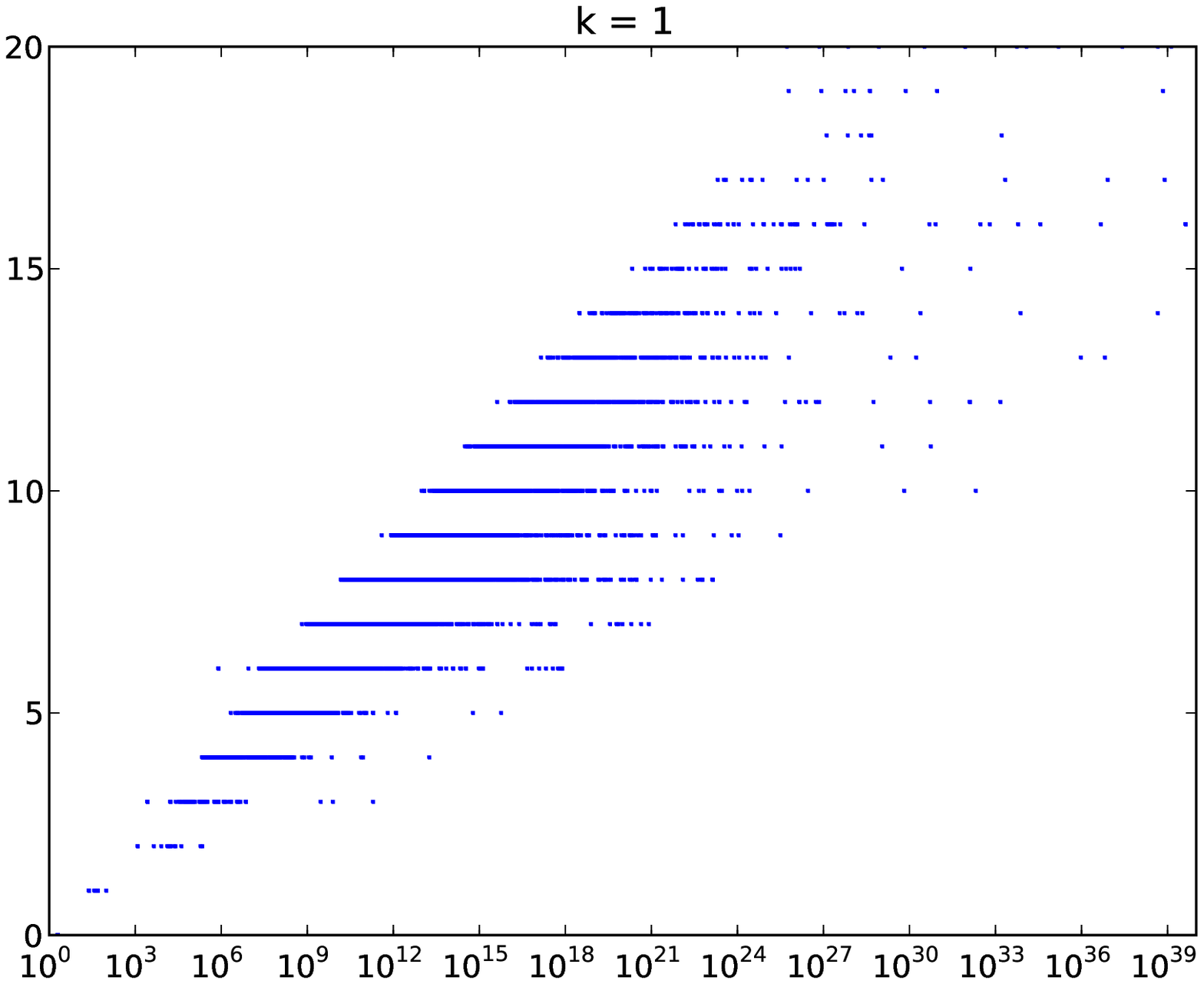}\includegraphics[width=0.5\columnwidth]{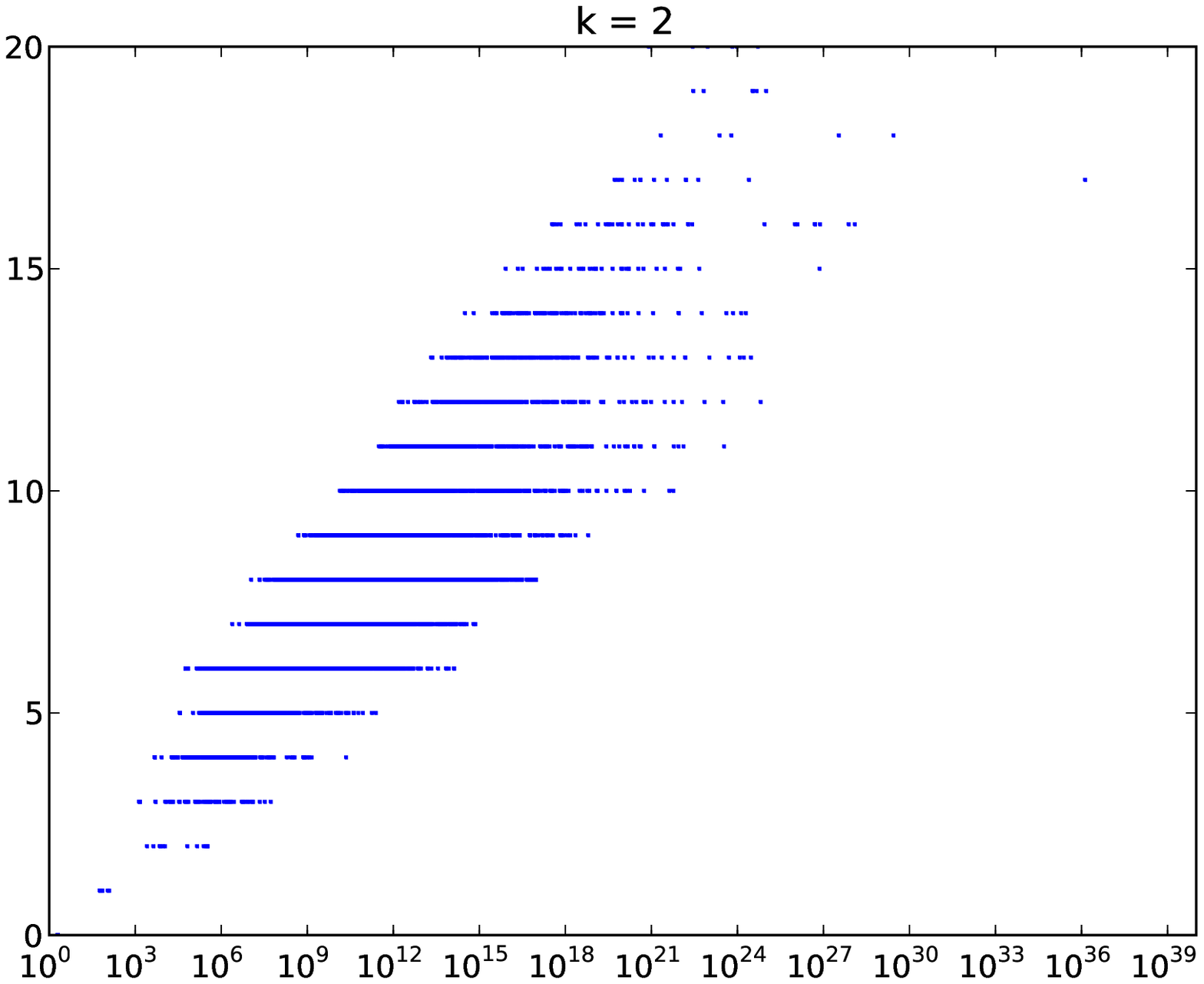}
\par\end{centering}

\begin{centering}
\includegraphics[width=0.5\columnwidth]{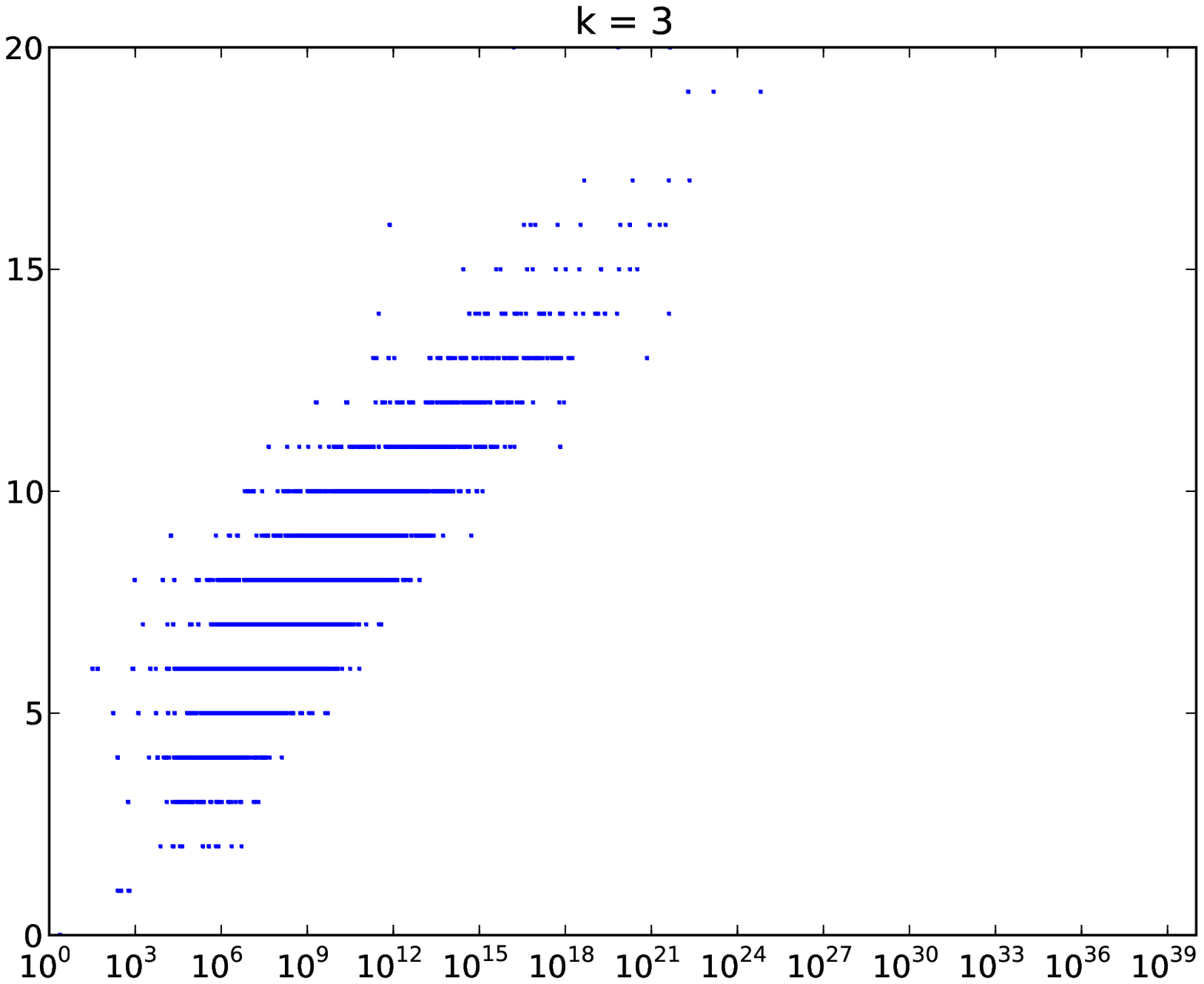}\includegraphics[width=0.5\columnwidth]{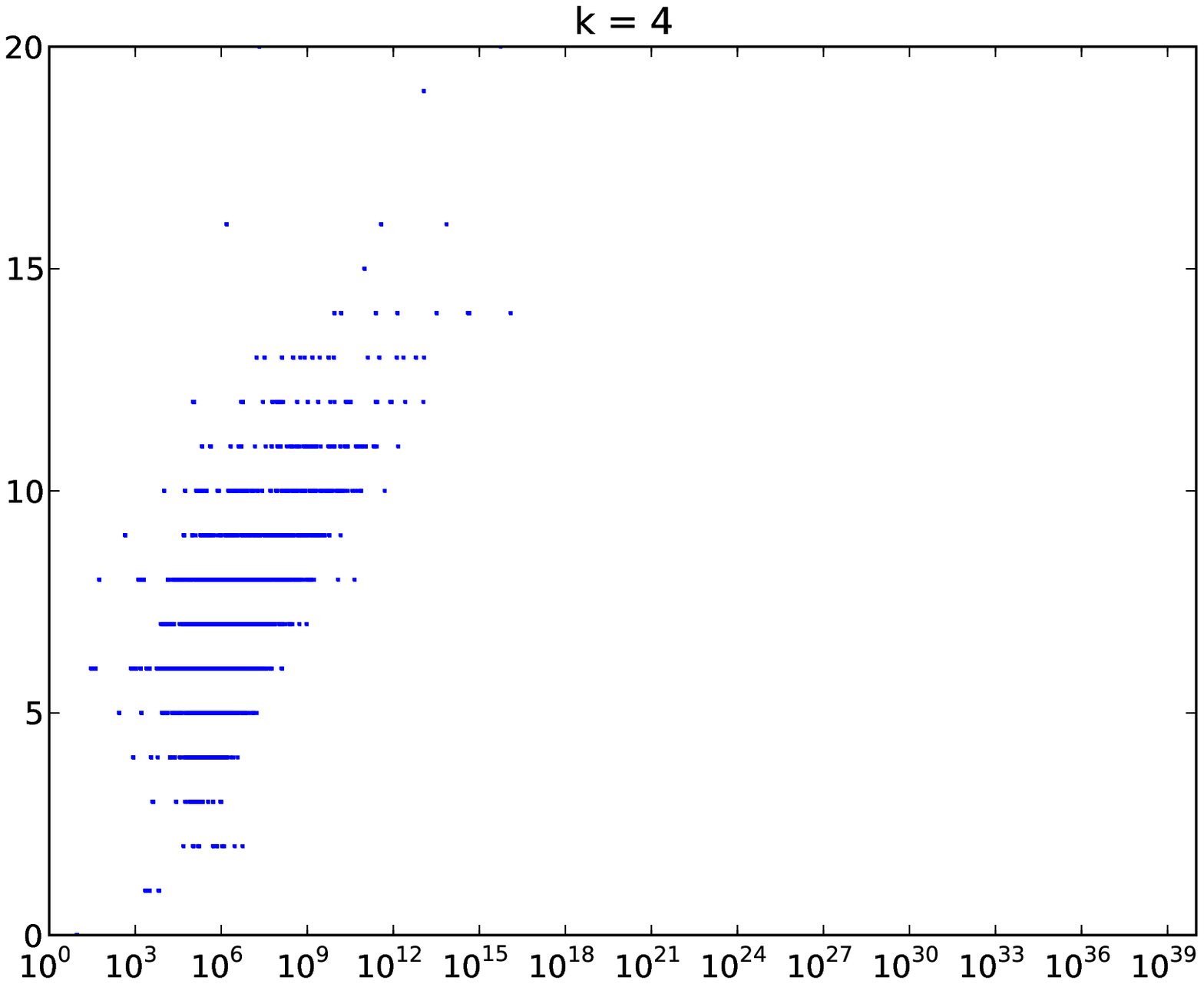}
\par\end{centering}

\caption{\label{fig:Password-length-vs.}Scatter plots of password length (Y
axis) versus strength (associated search space size on X axis).}

\end{figure}

In Figure \ref{fig:Password-length-vs.}, we highlight the relationship
between a password length and its strength. As the graphs show, the
assumption that longer passwords are stronger can only be regarded
as a rule of thumb: a short password containing infrequent characters
and/or sequences thereof can be actually stronger than a noticeably
longer one. The correlation between length and strength becomes weaker
as the $k$ parameter grows: long but weak passwords may be based
on predictable long patterns that are less efficiently predicted by
models based on lower $k$ values. For example, it is quite likely
that the {}``\texttt{abcd}'' sequence is followed by a {}``\texttt{e}'';
a model based on digraphs, though, cannot capture this and can only
model which character is more likely to follow a {}``\texttt{d}''.

\paragraph{Training Sets}

\begin{figure}
\begin{centering}
\includegraphics[width=1\columnwidth]{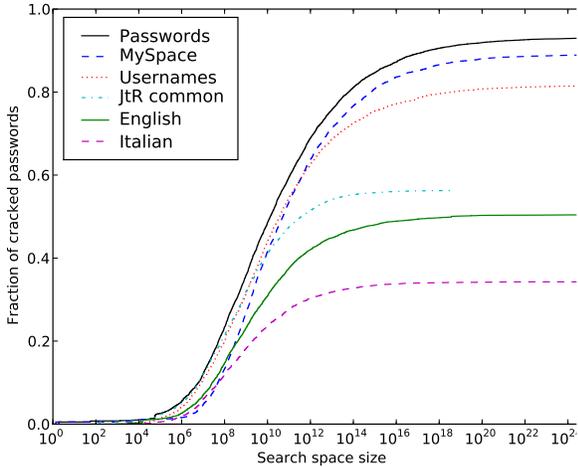}
\par\end{centering}

\caption{\label{fig:Comparing-training-sets.}Comparison of various training
sets for guessing passwords in our dataset $\left(k=2\right)$.}

\end{figure}

Figure \ref{fig:Comparing-training-sets.} illustrates how the choice
of training sets affects the attack performance. The training sets
used are our sets of usernames and passwords, the MySpace leaked passwords,
the JtR common password dictionary, and Italian and English dictionaries.

The most effective training set is the real password set. The {}``common
passwords'' dictionary from\emph{ }JtR is more representative of
real passwords than standard dictionaries, since it contains combinations
of characters, such as punctuation and digits, that don't appear in
standard dictionaries. Still, it appears that {}``average'' passwords
do not closely resemble the most common ones.

The case of MySpace passwords as training set is interesting: they
are close to the performance of our password dataset for strong passwords,
but they do not represent weak ones well. We believe this is due to
the over-representation of non-alphabetic characters, which are required
to be present in MySpace passwords. The difference in coverage on
strong passwords (around 5\% with equivalent search space size) can
also be attributed to this feature, as well as to the following factors:
\begin{itemize}
\item Difference in computer literacy: the MySpace sample contains only
the victims of a phishing attack;
\item Difference in language: MySpace users are distributed worldwide.
\end{itemize}
If a representative training set of real passwords is not available
to the attacker, usernames are by far the most effective training
set. It appears that, when users are asked to provide a username and
a password, they employ similar criteria. This is quite surprising
since the two strings need to satisfy very different, and arguably
conflicting, criteria: good usernames are easily memorable, while
a strong password has to be as difficult to guess as possible.

\paragraph{Usernames}

\begin{figure}
\begin{centering}
\includegraphics[width=1\columnwidth]{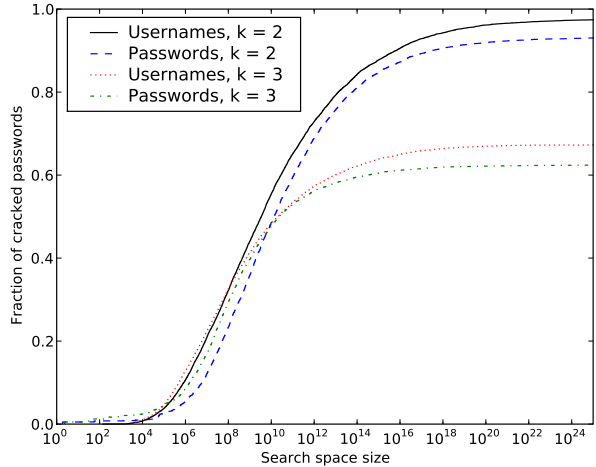}
\par\end{centering}

\caption{\label{fig:Search-space-for}Comparison of complexity between passwords
and usernames.}

\end{figure}

The former result suggests a consideration: usernames and passwords
are chosen simultaneously, when registering a new account. A user
wants both strings to be memorable, since the two are needed in order
to log on successfully. However, while there is no incentive in choosing
complex usernames, a security conscious user will commit some effort
to make his password more complex.

The difference in complexity between usernames and passwords is therefore
a way to measure the effort that users willingly put in making their
passwords more complex: while usernames can be very long or difficult
to guess, this is not likely to happen as the result of a conscious
attempt to do so.

In Figure \ref{fig:Search-space-for}, we compare the search space
size associated to usernames and passwords. Matching what we have
done with passwords, the training set used to guess a given username
consists of all the usernames except the one under scrutiny. It turns
out that the effort that users put in creating complex passwords is
measurable, but it is overall quite weak: given a choice for $k$
and a search space size, the percentage of {}``cracked'' usernames
never exceeds the cracked passwords by more than 15\%.

\section{Combined Strategy}

Our results confirm that no single strategy or technique is more effective
in reducing the search space: dictionaries are most effective in discovering
the weakest passwords; the coverage (fraction of passwords that are
in the dictionary) grows as the dictionary size grows, but this entails
a loss in precision (fraction of dictionary items that are actual
passwords). The Markov-chain based technique should be used when dictionaries
are exhausted. Higher values of $k$ obtain better results at first,
but after a number of attempts they become quite ineffective.

No single strategy is the best one for all cases; this, in fact, validates
the approach taken by password recovery systems that adopt bigger
and bigger dictionaries in cascade, and resort afterwards to Markov-based
techniques. In this section, we summarize our results by presenting
the results that an attacker would be able to obtain by using such
a technique.

Consistently with our approach of estimating the capabilities of the
attacker by excess in the face of uncertainty, we assume that the
attacker has access to a password training set which is as effective
as the one we obtain from the clear text. Furthermore, we also assume
that the attacker is able to predict the effectiveness of techniques
that we measured in Sections \ref{sec:Dictionary-Attack} and \ref{sec:Markov-Chain-Based-Attack}.

Based on this knowledge, using the training set as a dictionary, the
strategy for the dictionary-based first part of the attack is as follows:
\begin{enumerate}
\item Try the username;
\item Try the common passwords dictionary;
\item Try all passwords in the training set;
\item Try the English 1 dictionary;
\item Try the Italian 1 and 2 dictionaries;
\item Try the English 2, 3 and extra dictionaries;
\item Try all remaining JtR dictionaries;
\item Try mangling rules.
\end{enumerate}
If this approach is not sufficient, one should resort to the Markovian
model.

\begin{figure}
\begin{centering}
\includegraphics[width=1\columnwidth]{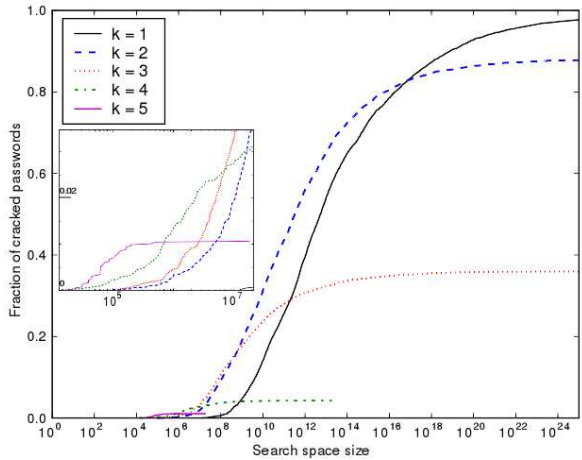}
\par\end{centering}

\caption{\label{fig:Variant-of-figure}Search space size for passwords that
are not found in any dictionary. In the inner frame, detail on the
first iterations.}

\end{figure}

In Figure \ref{fig:Variant-of-figure}, we show the search space for
the passwords that have not been discovered within any dictionary.
With respect to figure \ref{fig:Search-space}, there is a sharp decrease
in the success rate until the search space size reaches approximately
$10^{8}$. In particular, techniques with $k=5$ and $k=4$ are unsuccessful
to break more than, respectively, roughly 1\% and 4\% of the passwords.
This matches with the intuition that dictionary-based attacks are
more useful against the less complex passwords.

Based on the data represented in Figure \ref{fig:Variant-of-figure},
an efficient strategy for the attack would be as follows:
\begin{enumerate}
\item Try 500,000 candidates with the model based on $k=5$%
{};
\item Try 7,000,000 candidates with $k=4$%
{};
\item Try 700,000,000 candidates with $k=3$%
{};
\item Try $7\cdot10^{16}$ candidates with $k=2$ %
{};
\item Continue with $k=1$%
{}.
\end{enumerate}
\begin{table}
\begin{centering}
\begin{tabular}{|c|c|c|}
\hline 
Step & \#attempts & Cracked\tabularnewline
\hline
\hline 
Username & 1 & 2.88\%\tabularnewline
\hline 
Common passwords & 3,115 & 9.95\%\tabularnewline
\hline 
Training set & 10,431 & 28.83\%\tabularnewline
\hline 
English 1 & 36,574 & 30.51\%\tabularnewline
\hline 
Italian 1 & 98,511 & 32.25\%\tabularnewline
\hline 
Italian 2 & 373,834 & 36.31\%\tabularnewline
\hline 
English 2 & 632,613 & 37.18\%\tabularnewline
\hline 
English 3 & 722,215 & 37.69\%\tabularnewline
\hline 
English extra & 1,123,841 & 40.07\%\tabularnewline
\hline 
JtR - all dictionaries & 3,923,660 & 41.14\%\tabularnewline
\hline 
Mangling & 40,538,747 & 44.26\%\tabularnewline
\hline 
Markov chain - $k=5$ & 41,070,093 & 45.05\%\tabularnewline
\hline 
Markov chain - $k=4$ & 48,051,199 & 46.76\%\tabularnewline
\hline 
Markov chain - $k=3$ & \textasciitilde{}750,000,000 & 58.10\%\tabularnewline
\hline 
Markov chain - $k=2$ & \textasciitilde{}$7\cdot10^{16}$ & 91.06\%\tabularnewline
\hline 
Markov chain - $k=1$ & \textasciitilde{}$10^{40}$ & 99.71\%\tabularnewline
\hline
\end{tabular}
\par\end{centering}

\caption{\label{tab:Dictionary-based-multi-step-approach.}Cumulative number
of attempts and of guessed passwords for the multi-step approach.
Candidates that would be checked in more than one dictionary are counted
only once. For the Markov chain technique with $k\leq3$, the search
space has not been generated explicitly and its size has been approximated
with Algorithm \ref{alg:Implicit-approximation-of}.}

\end{table}

In Table \ref{tab:Dictionary-based-multi-step-approach.}, we summarize
the search space size and percentage of cracked passwords for each
of these steps. This is the answer to our original question: how many
attempts an attacker would need in order to guess a given percentage
of the passwords. By integrating this with system-specific knowledge
such as the computational cost needed to perform a single guess and
the amount of resources that the attacker has access to, it is possible
to estimate the percentage of passwords that are vulnerable to a given
attack.

\section{Conclusion}

As the bibliography of this work witnesses, the first studies on password
cracking date back to almost 30 years ago. Still, the techniques that
are used in state of the art password-cracking applications are quite
simple: decades of research suggest that it is possible to do better
than applying simple Markov chain-based modeling techniques.

The results of our measurement study may provide an explanation as
to why not much has been done in this direction: the diminishing returns
effect implies that, even if the size of the search space decreases
by orders of magnitude, the percentage of passwords that an attacker
would be able to crack in a given number of attempt would increase
only by a non-impressive percentage. In addition, it is likely that
an innovative strategy for exploring the search space would improve
over the state of the art only for a given interval of search space
sizes; the low-cost/high-reward part of the search space is already
easily covered by dictionaries of frequent passwords. When such an
attack proves ineffective, an attacker could change target to find
an easier prey, or use other means of attack which are not based on
the password strength, such as social engineering, phishing, or exploitation
of vulnerabilities in software or in the protocol: as the energies
instilled into an unsuccessful attack grow, the attack is more and
more likely to be unsuccessful in the future as well.

We focused on the strength of passwords chosen by users in the absence
of password strength enforcement. As pointed out in Section \ref{sec:Related-Work},
it is debatable that systems enforcing password complexity actually
increase security: they may instead lead users to circumvent the enforcement
techniques by adopting insecure behavior. To assess this, measuring
password complexity with and without enforcement should be coupled
with an analysis of user behavior.

Another interesting question yet to be addressed regards the correlation
between password strength and the domain they are related to. In particular,
how will the password strength of a user vary if getting an account
compromised would result in a noticeable loss? In \cite{Florencio2007Largescale},
some evidence that users actually choose better passwords for accounts
related to valuable assets (e.g., PayPal) is reported. Unfortunately,
the bit-strength measure adopted is quite simple. Further investigations
would be required to obtain actual figures in terms of attacker costs
in order to break an account.

\section{Acknowledgements}

The authors would like to thank Sebastian Porst and Roger Grimes for
having shared the set of MySpace passwords, and Cynthia Kuo, Sasha
Romanosky, and Lorrie F. Cranor for having shared their mnemonics
dictionary.

\bibliographystyle{abbrv}
\bibliography{matteodellamico}

\end{document}